\documentclass[]{aa}%
\usepackage{graphicx}
\usepackage{txfonts}
\begin{document}
\title{The nearby eclipsing stellar system $\delta$ Velorum}
\subtitle{II. First reliable orbit for the eclipsing pair\thanks{Reduced \'echelle spectra
of Delta Velorum are available at the CDS via anonymous ftp to cdsarc.u-strasbg.fr (130.79.128.5)
or via http://cdsweb.u-strasbg.fr/cgi-bin/qcat?J/A+A/
}}
\titlerunning{Eclipsing system $\delta$ Velorum}
\authorrunning{T. Pribulla et al.}


   \author{T. Pribulla \inst{1,2}, A. Merand \inst{3},  P. Kervella \inst{4}, M. Va\v{n}ko \inst{2},
           I.R. Stevens \inst{5}, R. Chini \inst{6,7}, V. Hoffmeister \inst{6}, O. Stahl \inst{8},
           A. Berndt \inst{1}, M. Mugrauer \inst{1}, \and M. Ammler-von Eiff \inst{9}}

   \institute{Astrophysikalisches Institut und Universit\"ats-Sternwarte,
              Schillerg\"a{\ss}chen 2-3, 07745 Jena, Germany\\
              \email{[alex,markus]@astro.uni-jena.de}
           \and
              Astronomical Institute, Slovak Academy of Sciences, 059~60
              Tatransk\'a Lomnica, Slovakia\\
             \email{[pribulla,vanko]@ta3.sk}
            \and
              European Southern Observatory, Alonso de Cordova 3107, Vitacura, Santiago,
              Chile\\
             \email{amerand@eso.org}
           \and
              LESIA, Observatoire de Paris, CNRSUMR8109, UPMC, Universit\'e Paris Diderot,
              5 Place Jules Janssen, 92195 Meudon, France\\
             \email{Pierre.Kervella@obspm.fr}
           \and
             School of Physics and Astronomy, University of Birmingham,
              Edgbaston, Birmingham, B15 2TT, United Kingdom\\
              \email{irs@star.sr.bham.ac.uk}
           \and
              Astronomisches Institut, Ruhr-Universit\"at Bochum, Universit\"atsstr. 150,
              44801 Bochum, Germany\\
              \email{[chini,vhoff]@astro.rub.de}
           \and
              Facultad de Ciencias, Universidad Cat\'{o}lica del Norte, Antofagasta, Chile
           \and
              ZAH, Landessternwarte K\"onigstuhl, 69117 Heidelberg, Germany\\
             \email{o.stahl@lsw.uni-heidelberg.de}
           \and
              Institut f\"ur Astrophysik, Georg-August-Universit\"at, Friedrich-Hund-Platz 1,
              37077 G\"ottingen, Germany\\
             \email{ammler@mps.mpg.de}
   }

   \date{Received September 15, 2010; accepted September 16, 2010}


  \abstract
   {The nearby multiple system $\delta$ Velorum contains a widely detached eclipsing binary and a third
   component. }
   {We take advantage of this system offering the opportunity to determine the set of fundamental
    parameters (masses, luminosities, and radii) of three coeval stars with sufficient precision
    to test models of stellar evolution.}
   {Extensive high-resolution spectroscopy is analyzed by the broadening function technique to
    provide the first spectroscopic orbit of the eclipsing pair. Simultaneous analysis of the
    spectroscopic data and the SMEI satellite light curve is performed to provide astrophysical
    parameters for the components. We use a modified Roche model assuming an eccentric orbit and
    asynchronous rotation.}
   {The observations show that components of the eclipsing pair rotate at about two-thirds of the
    break-up velocity, which excludes any chemical peculiarity and results in a non-uniform surface
    brightness. Although the inner orbit is eccentric, no apsidal motion is seen during the SMEI
    photometric observations. For the inner orbit, the orbital parameters are eccentricity
    $e$ = 0.290, longitude of the periastron passage $\omega = 109$\degr, and inclination
    89.0\degr.}
    {The component masses of $M_{\rm Aa} = 2.53\pm0.11$ M$_\odot$, $M_{\rm Ab} = 2.37\pm0.10$ M$_\odot$,
    and $M_B \sim 1.5$ M$_\odot$ combined with the inferred radii of the Aa and Ab components indicate
    that the eclipsing pair has already left the main sequence and that the estimated age of the system
    is about 400 Myr.
    }
   \keywords{stars: individual: HD 74956 ($\delta $ Vel) - stars: binaries: eclipsing -
    methods: observational - techniques: spectroscopic}
   \maketitle

\section{Introduction}
\label{intro}

$\delta$ Velorum is one of the fifty brightest stars in the sky ($V$ = 1.96) and is located
only 24.4 pc from the Sun ($\pi$  = 40.90$\pm$0.38 mas; ESA \cite{esa1997}). For a long time, it has
been known that it is a visual binary composed of the brighter component A with $H_P$ =1.99 and
the fainter component B with $H_P$ = 5.57 orbiting in a wide 142-year orbit (see Argyle et al.,
\cite{arg2002}).

Surprisingly, observations by visual observers, supported by photometry from the Galileo star
tracker, led to the discovery that the brighter component of the visual pair is an eclipsing binary
with $P$ = 45.15 days (Otero et al., \cite{oter2000}). No complete and reliable observations of
both eclipses, however, exist.

Spectroscopic observations of $\delta$ Velorum are also very limited. Levato (\cite{leva1972})
obtained medium-dispersion (40 \AA/mm) photographic spectra of the system and determined an
A1V spectral type and a projected rotational speed of $v \sin i$ = 85 km~s$^{-1}$.
$\delta$ Velorum was later included in the survey of early-type Hipparcos targets of Royer et al.
(\cite{roy2002}), who took one \'echelle spectrum of the system. The Fourier transforms of two
spectral lines were used to infer that $v \sin i$ = 150 km~s$^{-1}$, which is rather inconsistent
with the previous result. These measurements very probably included both A and B visual components.
No line doubling has been noticed.

The $\delta$ Velorum system was observed interferometrically by Kellerer et al. (\cite{kell2007}).
Their observations consisted of 17 squared visibility measurements using VLTI/VINCI. Although VLTI
clearly resolved the eclipsing pair, the presence of the visual component and the
small number of observations
caused the estimated parameters to be quite uncertain:
semi-major axis $a = 5.7\pm0.3~10^{10}$ m, radii of the components $R_{\rm Aa} = 6.0 \pm 0.5 R_\odot$,
$R_{\rm Ab} = 3.3\pm0.6 R_\odot$, eccentricity $e$ = 0.230$\pm$0.005, angle between the major axis and
line of sight $\omega' = -(20 \pm 3) \degr$\footnote{This angle transforms into a longitude of periastron
of $\omega = 90\degr - \omega'$.}, and longitude of the ascending node $\Omega = 27.4 \pm 1.2 \degr$
with a reduced $\chi^2_r$ = 2.6.

G\'asp\'ar et al. (\cite{gasp2008}) discovered a spectacular IR
bowshock around $\delta$ Velorum at 24 and 70$\mu$m using Spitzer/MIPS images. This very large
structure, $\sim$ 1\arcmin, was explained by the authors as a result of the heating and compression
of the interstellar medium by the photons from $\delta$ Velorum as the trio moves through the
interstellar medium. Kervella et al. (\cite{kerv2009}) resolved the wide visual pair
AB using the VISIR and NACO instruments at the VLT and obtained independent photometry of each
of the Aa, Ab, and B components. Their photometry did not infer the large radii of the components,
found by Kellerer et al. (\cite{kell2007}). Mid-infrared observations presented by the authors
exclude the presence of a circumstellar thermal excess around the system.

In spite of the above results, it is clear that a robust light-curve (hereafter LC)
and spectroscopic analysis are necessary to provide reliable orbital elements and
absolute parameters of the components.

\section{New observations}
\subsection{SMEI satellite photometry}
\label{smei}

Because of its high brightness, $\delta$ Velorum is a difficult target for ground-based
photometry. Unlike faint stars whose observation quality is dictated mostly by the shot
and read-out noise, observations of the brightest stars are affected by the lack of nearby and
sufficiently bright comparisons. Accompanying changes/differences in atmosphere transparency
result in substantial amount of red noise. Covering eclipses of the eclipsing pair from ground-based observations
is complicated by the long orbital period of $P$ = 45.15 days.

In the case of $\delta$ Velorum, the Solar Mass Ejection Imager (SMEI) provides a LC of
sufficient precision. In addition to its primary task, SMEI, attached to the Coriolis satellite,
is capable of producing high-precision photometric time-series for stars up to $V$ = 7
(see e.g., Bruntt et al., \cite{brun2006}; Tarrant et al., \cite{tarr2008};
Spreckley \& Stevens, \cite{spst2008}). In a similar way to
the MOST satellite (see Walker et al., \cite{most2003}), Coriolis stays close to the
dawn-dusk terminator with an orbital frequency of 14.17 cycles/day. Its three cameras take
narrow scans of the sky, but when combined, they provide almost full-sky coverage.
Of the 3 cameras that make up SMEI, camera three operates at a higher temperature and has
degraded photometric performance. We do not use data from camera 3 and this means
we do not have continuous coverage of $\delta$ Velorum. We refer to Goss et al. (\cite{goss2010})
and references therein, for more details about SMEI.

\begin{figure}
\centering
\includegraphics[width=8cm,clip]{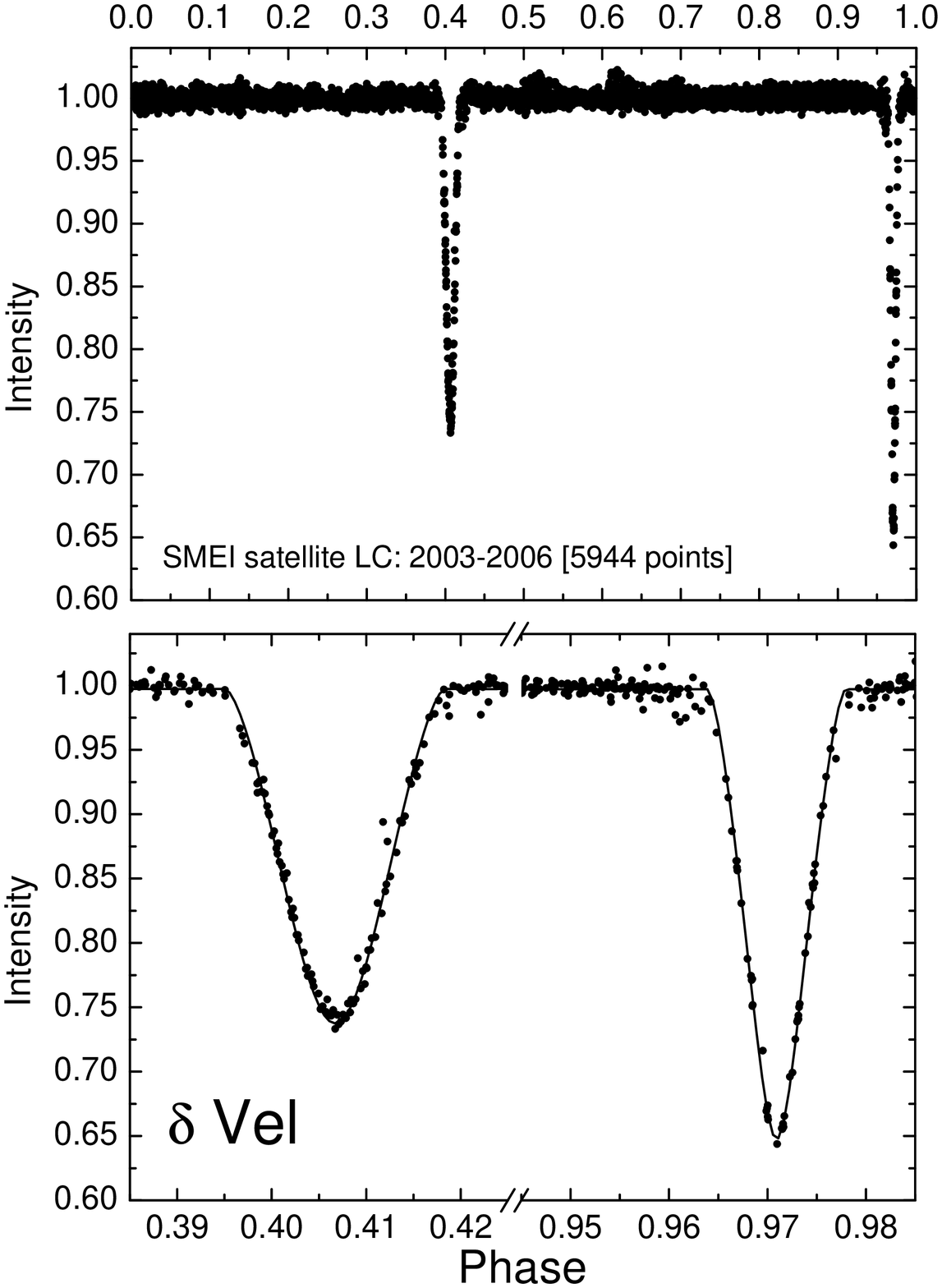}
\caption{Detrended and filtered (removing obvious outliers) SMEI LC of $\delta$ Vel.
         The data were phased using the following ephemeris for the periastron passage:
         To = HJD 2\,452\,528.950 + 45.15023$\times E$ (see Table~\ref{elements}).
         Only data from 2003-2006 are plotted.}
\label{smei_fig}
\end{figure}

The LC of $\delta$ Velorum generated from the SMEI images consist of 11195 points
covering 5.6 years. The data were obtained during seven observing seasons (winter
half of the year). The photometry is rather sparse, consisting of mostly one observation per
orbit. Owing to the very long orbital period of the eclipsing pair, and the short duration
of the eclipses, $\Delta T$ (pri) = 0.613 days and $\Delta T$ (sec) = 0.896 days,
only about 350 observations were obtained during the eclipses. By combining nine secondary minima
determined from the SMEI photometry and that obtained by the Galileo satellite tracker
(Otero, \cite{oter2000}) results in an ephemeris of Min~II = HJD 2\,447\,851.693(9) +
45.15023(7)$\times E$. The SMEI LC does not show apsidal motion. Because of the low angular
resolution of the SMEI cameras, both components of the visual pair contribute to the
extracted LC.

The quality of the SMEI photometry has been deteriorating since the launch of the satellite.
Hence, in the analysis presented here only earlier data (2003-2006) have been used
(Fig.~\ref{smei_fig}).

\subsection{BESO spectroscopy}
\label{beso}

Available spectroscopy of $\delta$ Velorum is very limited in spite of its brightness of
$V$ = 1.96. Two high-resolution \'echelle spectra are available in the ESO/FEROS data
archive taken on November 25, 2004 and January 7, 2009 (both outside eclipses). There is
practically no difference in the shape of the line profiles between the two spectra nor any
indication that $\delta$ Velorum is a SB2 system. The system has also been observed by ESO/HARPS.
Unfortunately, the spectra available in the ESO archive do not sufficiently cover the
orbital cycle and are of low signal-to-noise ratio (S/N).

\begin{figure}
\centering
\includegraphics[width=8cm,clip]{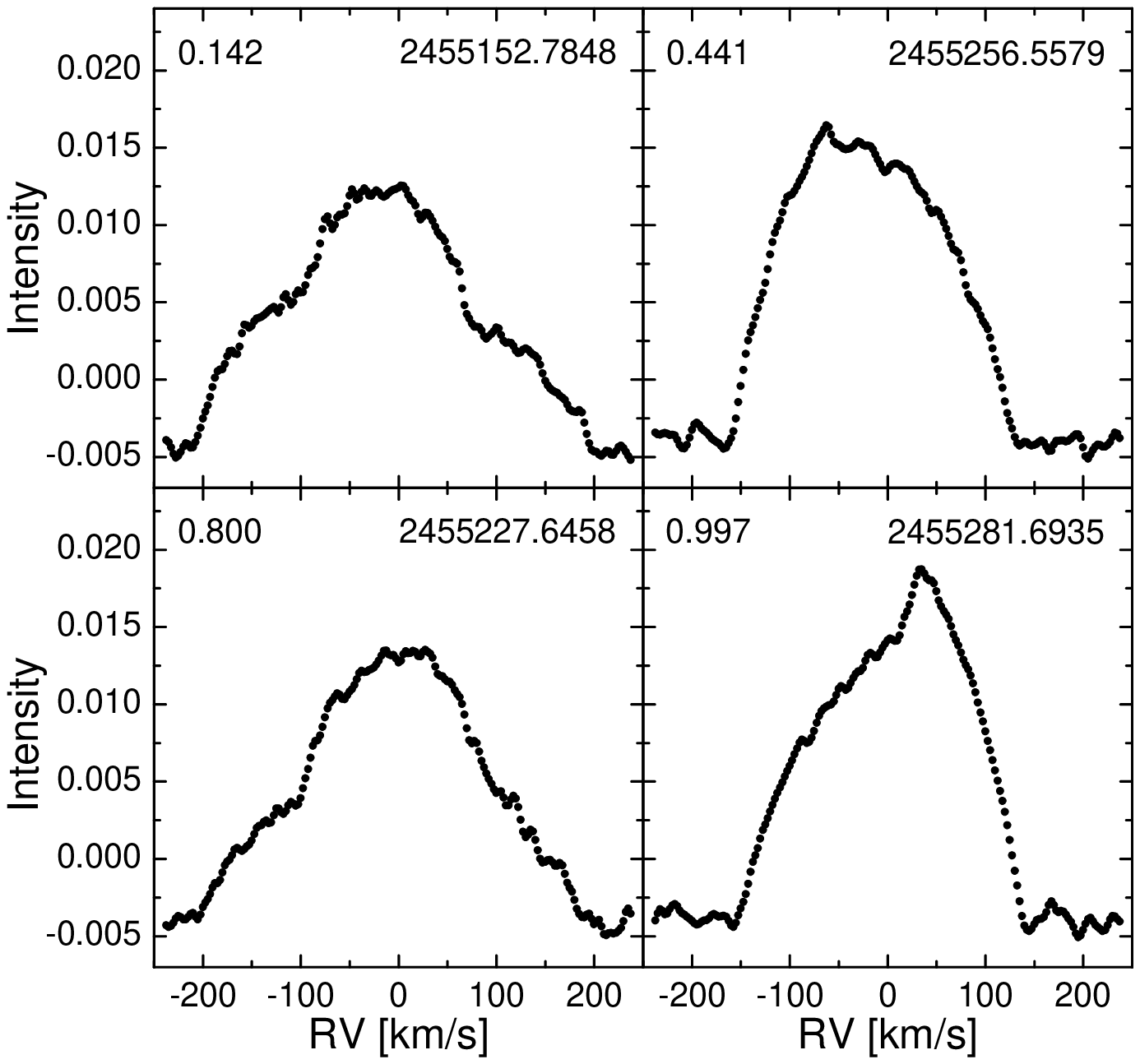}
\caption{Broadening functions close to the principal phases (counted from the primary eclipse,
         with the secondary occurring at phase 0.436): 0.142 (maximum separation of the components),
         0.441 (end of the secondary eclipse), 0.800 (maximum separation of the components), and
         0.997 (beginning of the primary eclipse). The radial velocity system is barycentric.
         Full mosaic showing all BFs including best fits is available only online.}
\label{bf_sel}
\end{figure}

New optical spectroscopy of $\delta$ Velorum has been obtained at the Cerro Armazones
Observatory using the BESO \'echelle spectrograph fiber-fed from the Cassegrain focus of the
1.5m Hexapod Telescope (hereafter HPT, see Fuhrmann et al., \cite{fuhr2010}). Sixty-three spectra
were obtained between April 2009 and April 2010. The spectra cover a $3530 - 8860 \AA$ wavelength
range. The data were
reduced using dedicated ESO-MIDAS scripts. The photometric reduction
includes overscan, bias, and flat-field correction. In subsequent steps individual \'echelle orders
were then extracted, wavelength calibrated, and normalized to the continuum. Finally, cosmic-ray spikes
were removed. Wavelength calibration was later improved using telluric bands close to 6900
and 7600$\AA$ using the spectrum of $\eta$ CMa as the telluric template. The improved RV system is
stable to about 100 m~s$^{-1}$ as indicated by the RV difference between the spectral bands
mentioned above. The absolute zero point of the system has not, however, been checked. Therefore, the
systemic velocity of $\delta$ Velorum may be offset from its true value.

\begin{table*}
\caption{Journal of spectroscopic observations at Cerro Armazones Observatory.
         \label{oca}
         }
\centering
\begin{tabular}{lcccr|lcccr}
\hline
\hline
Spectrum         &  BJD        & Phase  & S/N & Exp.& Spectrum         &  BJD        & Phase  & S/N & Exp.\\
                 & 2\,400\,000+&        &     & [s] &                  & 2\,400\,000+&        &     & [s] \\
\hline
20090429-01.fits & 54951.47364 & 0.6839 & 193 & 600 & 20100117-11.fits & 55214.72408 & 0.5145 & 233 & 600 \\
20091019-21.fits & 55124.88346 & 0.5247 & 195 & 300 & 20100120-06.fits & 55217.68788 & 0.5801 & 175 & 600 \\
20091019-22.fits & 55124.88728 & 0.5248 & 193 & 300 & 20100121-04.fits & 55218.66777 & 0.6018 & 259 & 600 \\
20091019-23.fits & 55124.89110 & 0.5248 & 198 & 300 & 20100122-05.fits & 55219.66618 & 0.6239 & 232 & 600 \\
20091020-20.fits & 55125.89151 & 0.5470 & 190 & 300 & 20100123-06.fits & 55220.62644 & 0.6452 & 223 & 600 \\
20091020-21.fits & 55125.89535 & 0.5471 & 196 & 300 & 20100125-05.fits & 55222.68527 & 0.6908 & 207 & 600 \\
20091021-18.fits & 55126.89139 & 0.5691 & 203 & 300 & 20100127-04.fits & 55224.66181 & 0.7346 & 204 & 600 \\
20091022-25.fits & 55127.87589 & 0.5909 & 207 & 300 & 20100128-02.fits & 55225.67460 & 0.7570 & 229 & 600 \\
20091022-26.fits & 55127.88004 & 0.5910 & 203 & 300 & 20100130-02.fits & 55227.64583 & 0.8007 & 176 & 600 \\
20091022-27.fits & 55127.88384 & 0.5911 & 215 & 300 & 20100131-02.fits & 55228.70596 & 0.8242 & 168 & 600 \\
20091023-18.fits & 55128.88626 & 0.6133 & 218 & 300 & 20100208-01.fits & 55236.65951 & 0.0003 & 113 & 600 \\
20091023-19.fits & 55128.89015 & 0.6134 & 193 & 300 & 20100208-06.fits & 55236.76622 & 0.0027 & 121 & 600 \\
20091024-20.fits & 55129.88268 & 0.6354 & 188 & 300 & 20100211-04.fits & 55239.66155 & 0.0668 & 272 & 600 \\
20091112-15.fits & 55148.81797 & 0.0548 & 212 & 300 & 20100214-10.fits & 55242.71579 & 0.1345 & 210 & 600 \\
20091112-16.fits & 55148.82182 & 0.0549 & 234 & 300 & 20100217-09.fits & 55245.66592 & 0.1998 & 241 & 600 \\
20091116-05.fits & 55152.78108 & 0.1426 & 247 & 300 & 20100220-07.fits & 55248.58518 & 0.2644 & 202 & 600 \\
20091116-06.fits & 55152.78487 & 0.1426 & 261 & 300 & 20100226-08.fits & 55254.68385 & 0.3995 & 383 &1800 \\
20091116-09.fits & 55152.85198 & 0.1441 & 221 & 300 & 20100228-04.fits & 55256.55795 & 0.4410 & 262 & 600 \\
20091118-04.fits & 55154.76307 & 0.1865 & 213 & 300 & 20100228-13.fits & 55256.72306 & 0.4447 & 281 & 600 \\
20091118-05.fits & 55154.76687 & 0.1865 & 224 & 300 & 20100304-07.fits & 55260.66669 & 0.5320 & 293 & 600 \\
20091121-07.fits & 55157.77206 & 0.2531 & 222 & 300 & 20100307-10.fits & 55263.64569 & 0.5980 & 296 & 600 \\
20091121-08.fits & 55157.77950 & 0.2533 & 234 & 300 & 20100310-09.fits & 55266.63989 & 0.6643 & 221 & 600 \\
20091130-02.fits & 55166.79663 & 0.4530 & 153 & 300 & 20100313-06.fits & 55269.63109 & 0.7306 & 215 & 600 \\
20091130-03.fits & 55166.80047 & 0.4531 & 160 & 300 & 20100316-08.fits & 55272.61929 & 0.7968 & 328 & 600 \\
20091209-08.fits & 55175.70075 & 0.6502 & 100 & 300 & 20100322-07.fits & 55278.62672 & 0.9298 & 265 & 600 \\
20091209-09.fits & 55175.70490 & 0.6503 & 100 & 300 & 20100325-07.fits & 55281.60965 & 0.9959 & 276 & 900 \\
20100109-09.fits & 55206.71990 & 0.3372 & 193 & 300 & 20100325-10.fits & 55281.69350 & 0.9977 & 287 & 900 \\
20100109-10.fits & 55206.72635 & 0.3373 & 205 & 300 & 20100326-06.fits & 55282.57737 & 0.0173 & 333 & 900 \\
20100111-12.fits & 55208.76993 & 0.3826 & 163 & 600 & 20100326-10.fits & 55282.68869 & 0.0198 & 314 & 900 \\
20100112-09.fits & 55209.75303 & 0.4044 & 158 & 600 & 20100328-09.fits & 55284.60675 & 0.0623 & 400 & 600 \\
20100114-07.fits & 55211.69796 & 0.4475 & 167 & 600 & 20100331-06.fits & 55287.57912 & 0.1281 & 274 & 600 \\
20100116-07.fits & 55213.69359 & 0.4917 & 251 & 300 &                  &             &        &     &     \\
\hline
\end{tabular}
\tablefoot{The filename contains evening date
         of observation encoded as yyyymmdd. Barycentric Julian
         dates of mid-exposures are given. The S/N ratio was estimated from
         line-free continuum close to the Mg II 4481 line. Phases
         of mid-exposures were computed using the following ephemeris
         for the primary minimum: 2\,447\,832.0075 + 45.15023$\times E$.}
\end{table*}

According to the VLT/NACO imaging of Kervella et al. (\cite{kerv2009}), the separation of
the optical pair A-B was about 0.6\arcsec\ in 2008. Because of a PSF of ~3-5" at the HPT, both
visual components were included in the fiber entrance. Hence, it is reasonable to assume that
no light from the faint companion has got lost. The PSF required exposure times of typically
300-900 seconds, resulting in S/N ratios ranging from about 100 to 400. The journal of
observations can be found in Table~\ref{oca}.

The aforementioned sixty-three spectra cover more-or-less uniformly the orbital cycle of $\delta$ Velorum. Several
spectra were intentionally taken during the eclipses. All reduced one-dimensional spectra
will be available at the CDS.

\section{Data analysis}
\subsection{Orientation of the orbit}
\label{orientation}

\begin{figure}
\centering
\includegraphics[width=8cm,clip]{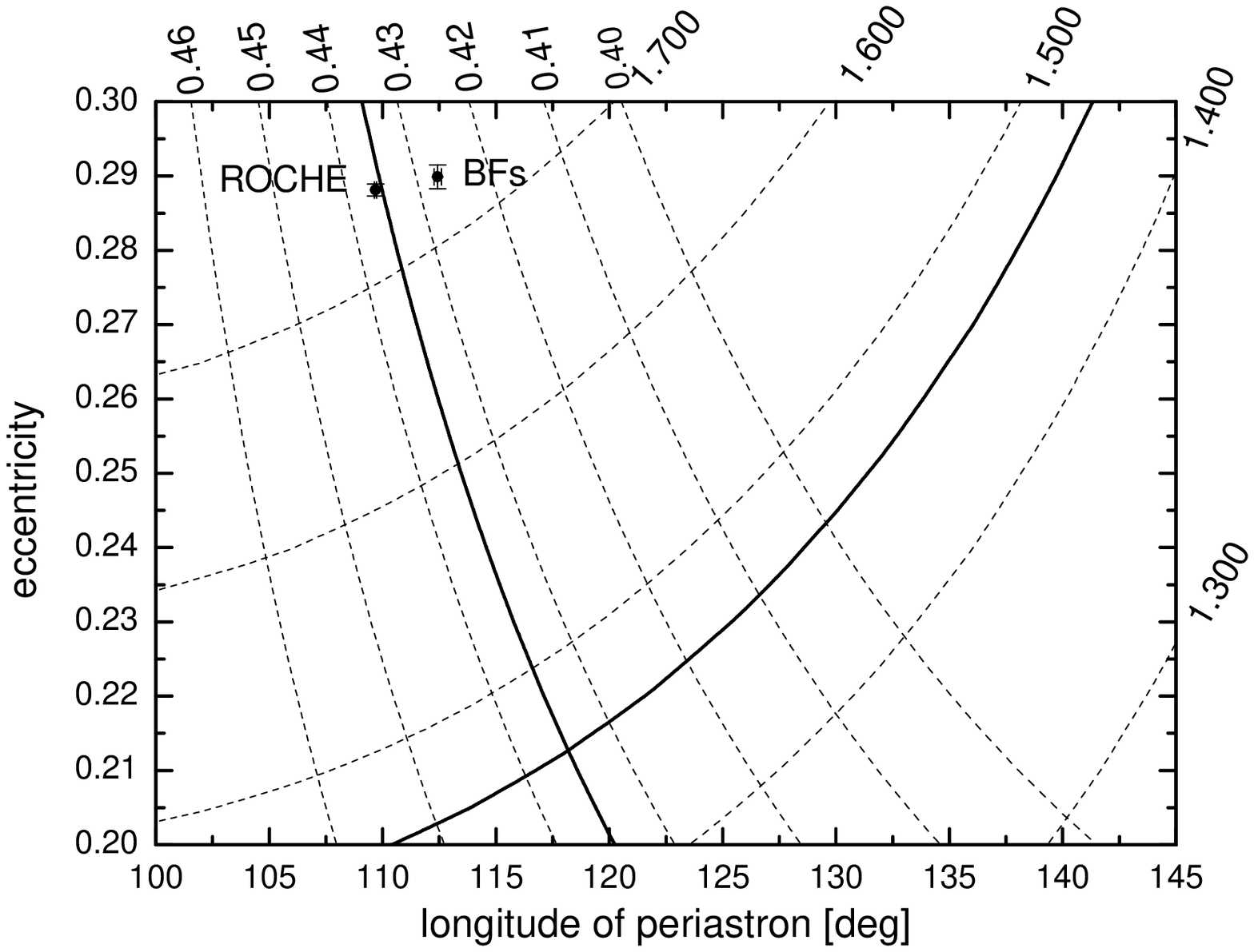}
\caption{Phase of the secondary minimum (labelled values are $0.40 - 0.46$; the observed phase is 0.436)
         and ratio of minima durations (with labelled values $1.3 - 1.7$;
         the observed ratio is 1.462). The lines are valid for an inclination angle of $\sim 90$\degr.
         Thick solid lines represent the observed values and their intersection indicates that $e = 0.20 - 0.22$
         and $\omega = 115 - 120\degr$. The resulting orientation and eccentricity of the orbit plotted for simple BF
         modelling (Section~\ref{doppler}) and ROCHE modelling (Section~\ref{simultan}) shows that $i < 90$\degr.}
\label{orient}
\end{figure}

In principle, the orientation and eccentricity of the orbit can be estimated from the position (phase) of the
secondary minimum and ratio of the minima durations (see analysis of Kellerer et al., \cite{kell2007}).
If we denote the deeper minimum as the primary and assume that the orbit is seen edge-on, from the
observed minima durations and secondary minimum
phase in the case of $\delta$ Velorum, one infers that $e = 0.20 - 0.22$ and $\omega = 115 - 120\degr$
(see Fig.~\ref{orient}). The longitude of the periastron refers to the orbit of the {\it component
eclipsed in the primary minimum}. When we refer to the orbit of the component eclipsed
in the secondary minimum, we must convert $\omega \rightarrow \omega + 180\degr$.

The case of eccentric orbits with $i \neq 90\degr$ is not that simple because (i) the ratio of minima depths
does not reflect the ratio of the surface brightnesses of the components (in the deeper, primary minimum even
the {\it cooler} and {\it less massive} component can be eclipsed), (ii) the ratio of minima width cannot be
used to find $e \sin \omega$. If $\omega \neq 90\degr$ nor $270\degr$, for sufficiently low inclination
angles the minimum occurring further away from the periastron can disappear and the observer can detect only
one system of minima. This is the case for e.g., BD+37~410 (Pribulla et al., \cite{prib2010}) or
NY~Cep (see Holmgren et al., \cite{holm1990}).

None of the minima of $\delta$ Velorum (see Fig.~\ref{smei_fig}) shows a
constant brightness interval. This means that at least one minimum is partial\footnote{Total minimum
can still be a transit, curved because of the limb darkening.} with $i < 90\degr$. Hence, exact eclipse
modeling is needed to derive reliable parameters.

\subsection{Extracting Doppler information and a preliminary orbit}
\label{doppler}

Besides the Balmer series, there are few other strong lines in the optical spectrum of
$\delta$ Velorum. The optimal spectral region for extracting Doppler information is the blue spectrum
between $4370 - 4605$ \AA. In addition to the fairly strong Mg II 4481, line it contains many weaker
metallic lines (mostly of Fe I) but is devoid of hydrogen lines. This blue part of the
spectra was deconvolved using a high-resolution synthetic spectrum corresponding to
$T^{\rm eff}$ = 9500~K, $\log g$ = 4.5, solar metallicity,  \footnote{The spectrum was taken
from the Pollux database available online at http://pollux.graal.univ-montp2.fr/.}
and, a non-rotating star to obtain broadening functions (BFs) using the formalism developed
by Rucinski (\cite{ruc1992}). Extracted BFs were smoothed by convolving with a Gaussian
function ($\sigma$ = 10 km~s$^{-1}$) to match the spectral resolution of BESO.

The template provided a very good match to the observed spectra indicated by the BF integral being
close to unity (on average about 0.97). Using a hotter template with $T^{\rm eff}$ = 10000~K,
$\log g$ = 4.5, and metallicity 0.5 dex lower than the solar one, we derived BF integrals of
about 1.58. The BF integral depends on the strength of metallic lines in the region, hence
a good spectral match can be provided by synthetic template of lower temperature and lower
metallicity or by using a template of higher temperature and higher metallicity.

Extracted BFs clearly show that $\delta$ Velorum is a SB2. Unfortunately, profiles of the
components are always blended (see Fig.~\ref{bf_sel}). The orbital motion can be, however,
well seen from strong changes in the total width and shapes of BFs.
The deformed shapes of BFs close to spectroscopic conjunctions confirm that eclipses are detected
in the system.
We note that the extracted BFs do not show any presence of the visual
component B in the system despite its expected contribution of about 3.7\%
(in the wide $H_p$ passband\footnote{The maximum response of the Hipparcos photometric system
lies within spectral range used to extract the BFs.}). We also produced an average spectrum
of $\delta$ Velorum and extracted the corresponding BF. In this way, the orbital motion of the
eclipsing pair is smeared-out but no additional component can be seen. This indicates
that the visual component B is also a rapid rotator or a SB1/SB2 system.
Another possible explanation is that the spectral type of the third component is
too different to be picked up in the spectrum using the A0 template.

Because the profiles of the components are never separated and we observe only blends of
variable shape, it is impossible to directly determine the RVs of the components. Hence, we
modeled the whole dataset (with a single set of parameters) assuming that (i) the width and
shape of the rotational profiles of the components do not change with phase, (ii) the relative
intensity of the components is constant, (iii) the limb darkening is the same for both
components, (iv) there is no apsidal motion visible in our data (as indicated by the long-period
of the eclipsing pair), and (v) stars rotate as solid bodies (no differential
rotation). To fulfill the first two conditions, the spectra taken during the eclipses were
neglected. The standard deviation of each BF was determined using its violet part which was
always outside of the component's profiles.

\begin{table}
\caption{Parameters of the eclipsing pair in the $\delta$ Velorum system.
         \label{elements}
         }
\centering
\begin{tabular}{lc|c|c}
\hline
\hline
                                              &             & BF fitting   & ROCHE        \\
{\bf Assumed parameters}                      &             &              &              \\
\hline
$P$                                           & [days]      & 45.15023(7)  & 45.15023(7)  \\
$\beta_{\rm Aa} = \beta_{\rm Ab}$             &             &   --         &  0.25        \\
$A_{\rm Aa} = A_{\rm Ab}$                     &             &   --         &  1.00        \\
$\pi$                                         & [mas]       &   --         & 40.90(38)    \\
$L_3 = L_{\rm B}/(L_{\rm Aa} + L_{\rm Ab})$   &             &   --         &  0.048       \\
$T^{\rm eff}_{\rm Aa}$                        & [K]         &   --         &  9420        \\
\hline
{\bf Optimized parameters}                    &             &              &              \\
\hline
$T_0$                                         & [HJD]       &   529.185(11)&  528.950(5)  \\
$e$                                           &             & 0.2899(16)   & 0.2881(8)    \\
$i$                                           & [deg]       &   --         &  89.00(3)    \\
$\omega$                                      & [deg]       &  112.43(16)  & 109.69(5)    \\
$\Omega_{\rm Aa}$                             &             &   --         &  34.3(4)     \\
$\Omega_{\rm Ab}$                             &             &   --         &  35.7(4)     \\
$F_{\rm Aa}$                                  &             &   --         &  43.2(7)     \\
$F_{\rm Ab}$                                  &             &   --         &  50.3(9)     \\
$q$ = $M_{\rm Ab}/M_{\rm Aa}$                 &             &   0.938(11)  &  0.937(10)   \\
$K_{\rm Aa} + K_{\rm Ab}$                     & [km/s]      &   104.8(2)   &  106.0(15)   \\
$V_A$                                         & [km/s]      &  $-12.07(5)$ &  $-10.4(5)$  \\
$v \sin i_{\rm Aa}$                           & [km/s]      &   139.48(10) &   139.4      \\
$v \sin i_{\rm Ab}$                           & [km/s]      &   146.84(13) &   146.5      \\
$\chi^2_r$ (LC)                               &             &    --        &   0.852      \\
$\chi^2_r$ (BF)                               &             &    2.710     &   1.945      \\
\hline
{\bf Computed parameters}                     &             &              &              \\
\hline
$A \sin i$                                    & [R$_\odot$] &   89.5(2)    &   90.6(9)    \\
$M_{\rm Aa} \sin^3 i$                         & [M$_\odot$] &   2.44(2)    &    2.53(11)  \\
$M_{\rm Ab} \sin^3 i$                         & [M$_\odot$] &   2.29(2)    &    2.37(10)  \\
$L_{\rm Aa}$                                  & [L$_\odot$] &     --       &   56.3(17)   \\
$L_{\rm Ab}$                                  & [L$_\odot$] &     --       &   47.1(25)   \\
$l_{\rm Ab}/l_{\rm Aa}$                       &             &   0.813(3)   &   0.822(4)   \\
$R_{\rm Aa}$                                  & [R$_\odot$] &     --       &    2.83(4)   \\
$R_{\rm Ab}$                                  & [R$_\odot$] &     --       &    2.54(5)   \\
$\log g_{\rm Aa}$                             &   [cgs]     &     --       &    3.90(2)   \\
$\log g_{\rm Ab}$                             &   [cgs]     &     --       &    3.97(3)   \\
\hline
\hline
\end{tabular}
\tablefoot{In both cases (rotational profile fitting to BFs, Section~\ref{doppler}, and full
         ROCHE modelling, Section~\ref{simultan}), the orbital period, $P$, corresponds to
         the best linear fit to the Galileo and SMEI secondary minima (see Section~\ref{smei}).
         Third light $L_3$ has been estimated for the SMEI passband using model atmospheres.
         Heliocentric Julian date of the periastron passage is $-2\,452\,000$.  Reduced
         $\chi^2_r$ is separately given for LCs and BFs. Generalized equipotentials are given
         for the mean distance (not in the periastron passage).
         Projected rotational velocities in the case of the combined ROCHE solution were not optimized
         but computed from other parameters. The ratio of fluxes, $l_{\rm Ab}/l_{\rm Aa}$, is given in the
         SMEI passband.}
\end{table}

In the case of solid-body rotation and a linear limb-darkening law, the rotational profile is
an analytic function. The BF observed outside the eclipses is just the sum of two rotational profiles
as given by Gray (\cite{gray1976}). The limb darkening coefficient was fixed to be $u_{\rm Aa} =
u_{\rm Ab}$ = 0.522 as appropriate for an A0V star ($T^{\rm eff}_{\rm A}$ = 9420~K, Popper,
\cite{popp1980}) with $\log g = 4.5$ and $\lambda$ = 4400 \AA. Global fitting to all observed BFs included
the following parameters: orbital eccentricity $e$; longitude and time of the periastron
passage $T_0$, $\omega$; systemic velocity $V_A$; sum of semi-amplitudes of the RV
changes $K_{\rm Aa} + K_{\rm Ab}$; mass ratio $q$ = $M_{\rm Ab}/M_{\rm Aa}$;
background level of BFs\footnote{For both a perfect match between the spectral types of the template and
observed star and proper rectification to the continuum, it should be zero.} $B_0$; intensities
of the profiles $I_{\rm Aa}$, $I_{\rm Ab}$; and projected rotational velocities of the components
$v_{\rm Aa} \sin i$, $v_{\rm Ab} \sin i$. The orbital period $P$ was held fixed at the photometrically
determined value, $P$ = 45.15023 days, because of the relatively short time-span of the observations.

The convergence process was repeated starting at many (dozens) parameter sets. Modeling showed
that the component eclipsed in the primary minimum is the more massive of the two.

The optimization always resulted in the same parameters (listed in Table~\ref{elements})
indicating uniqueness of the solution (see online Fig.~5 showing fits to all BFs outside
eclipses). The effect of the limb darkening is small but not negligible. For the acceptable
temperature range, 8750~K (A2V) to 10000~K (B9.5V), the principal parameters lie within the
following ranges: $e$ = [0.2889, 0.2924], $K_{\rm Aa} + K_{\rm Ab}$ =  [104.41, 105.27] km~s$^{-1}$,
$v_{\rm Aa} \sin i$ = [139.11, 140.21] km~s$^{-1}$, and $v_{\rm Ab} \sin i$ = [146.48, 147.58]
km~s$^{-1}$. It is clear that the systematic uncertainties (connected with the uncertain temperature
and the limb darkening coefficient) are larger than those
determined from residuals and the covariance matrix. The reduced $\chi^2_r$ corresponding to the tested
temperature range changes by only 3.1\% (lowest $\chi^2_r$ occurs for $T^{\rm eff}$ = 8750~K). For
any solution, it is clear that the component being eclipsed in the primary minimum is the more
massive of the two. The goodness of the fit, $\chi^2_r = 2.71$ is indicative of underestimated errors
in the BFs or/and a too simple model being used to fit the data.

For the solution corresponding to A0V stars, the ratio of fluxes is

\begin{equation}
\frac{l_{\rm Ab}}{l_{\rm Aa}} = \frac{I_{\rm Ab} v_{\rm Ab} \sin i}{I_{\rm Aa} v_{\rm Aa} \sin i} = 0.813(3).
\end{equation}

The brightness ratio should be interpreted with caution because in extracting BFs the same
template was used for both components. In the case that the spectral type of the secondary was
later (or better strength of the metallic lines was larger) the estimated light ratio is
the upper estimate. Results of the following section (\ref{simultan}) indicate, however, that the
temperatures of the components are very similar.

The corresponding masses of the components are $M_{\rm Aa} \sin^3 i$ = 2.439(20) M$_\odot$, and
$M_{\rm Ab} \sin^3 i$ = 2.288(18) M$_\odot$ (the mass ratio is then 0.938(11)). Considering the full
acceptable range of temperatures, the masses fall within $2.4308 < M_{\rm Aa} \sin^3 i < 2.4516$
M$_\odot$ and $2.2709 < M_{\rm Ab} \sin^3 i < 2.2950$ M$_\odot$. Because the inclination angle
is $\sim$ 89\degr (see Section~\ref{simultan}), the projected masses are close to the true
masses (a 1\% increase in mass occurs for an inclination $i$ = 85.3\degr).

The BFs, very probably, contain a weak signature of the visual component B. Its separate
spectrum would, however, be needed to remove its influence on BFs and the determined orbital elements.
If it were a single but rapidly-rotating star (having a profile with a constant RV), it would
effectively bring the components of the eclipsing pair together, reducing $(K_{\rm Aa} + K_{\rm Ab})$
and also alter the derived rotational velocities. Because of a severe blending of the primary
and secondary lines, it is also impossible to check whether the components rotate as a solid body or
differentially.

\subsection{Simultaneous analysis of photometry and spectroscopy}
\label{simultan}

Assuming that the apsidal motion is very slow\footnote{Using formulae of Hilditch
(\cite{hild2001}), resulting parameters from Table~\ref{elements} and corresponding
apsidal constants adopted from Claret (\cite{clar2004}) for 400 Myr stellar model with
overshooting (see Section \ref{triple}) leads to $U \sim 6$ Myr.} and the orbital period is stable,
we can combine SMEI (broadband) LC and OCA \'echelle spectroscopy (63 BFs) to help us derive
consistent parameters for the system.

Data modeling has been performed using an updated version of the code {\it ROCHE} (Pribulla,
\cite{prib2004}). The code assumes Roche model defining the surface geometry and local gravity
on the components. Solid-body rotation was assumed. It is assumed (as suggested by Wilson,
\cite{wils1979}) that the components
can adjust their shape\footnote{In the case of $\delta$ Velorum, an eccentric orbit causes very
small changes in the true radii with orbital phase, e.g., $\Delta R_{\rm point}/R \sim 7~10^{-5}$.
The shapes of the components are much more affected by fast asynchronous rotation.}
to slightly changing equipotential surfaces in the case of an eccentric orbit. The generalization
of the Roche model for asynchronous rotation was also applied. Surface grids are derived
from an icosahedron, resulting in practically equal elements. For both stars, it was assumed
that the Von Zeipel (\cite{zeip1924}) law (appropriate for radiative envelopes) dictates local temperature.
Single mutual reflection/irradiation was computed.

The optimized parameters are as follows: inclination angle $i$, generalized equipotentials
$\Omega_{\rm Aa}$, $\Omega_{\rm Ab}$, asynchronous rotation factors $F_{\rm Aa}$,
$F_{\rm Ab}$\footnote{For bound rotation, $F_{\rm Aa} = F_{\rm Ab} = 1$.}, mass ratio $q$,
polar temperature of the secondary $T^{\rm eff}_{\rm Ab}$, sum of semi-amplitudes $K_{\rm Aa}
+ K_{\rm Ab}$, systemic velocity $V_{\rm A}$, longitude of periastron $\omega$, eccentricity
$e$, global normalization factor of BFs (should be unity in the case of perfect template match),
and the global level of the BF background. Linear limb darkening coefficients $u_{\rm Aa}$, $u_{\rm Ab}$
were automatically recalculated, interpolating from tables of van Hamme (\cite{hamm1993})
according to the mean surface gravity and wavelength range (separately for LCs and BFs).
Local surface intensities were interpolated from tables of Lejeune et al. (\cite{leje1997})
for each surface grid point. Because of the changing distance and shape of the components,
both the surface grid and local parameters had to be recalculated for each step in phase
(360 steps/orbit). During the eclipses, a four-times finer phase step was used.

The scatter in {\it whole} individual datasets was estimated by performing polynomial fitting to
constant regions: in the case of the SMEI LC, we used out-of-eclipse phases, for BF parts
outside the blended profile. To constrain the solution more tightly, only eclipse
parts of the SMEI LC were used. Resulting parameters are listed in Table~\ref{elements}.
The combined solution gives true (not projected) masses of the components as
$M_{\rm Aa} = 2.53\pm 0.11$ M$_\odot$ and $M_{\rm Ab} = 2.37\pm 0.10$ M$_\odot$ ($\sin^3 i = 0.99954$).

The best fits to both eclipses, as observed by the SMEI satellite, are shown in Fig.~\ref{smei_fig}, and
fits to all BFs are available as online material only (online Fig.~6). The best fits to the BFs including
eclipses do not uncover any systematic discrepancies. Small differences in the shapes of
the minima branches of the SMEI LC can, however, be seen. This may be the result of a simplified
treatment of the limb darkening effect\footnote{The limb-darkening coefficients are assumed to
be constant over the surface of components.} or small departures of the component shapes from
the generalized Roche model. By comparing spectroscopic elements obtained by assuming two
limb-darkened rapidly rotating spheres (Section \ref{doppler}) to those obtained by the more
realistic modeling by assuming a Roche geometry including all proximity effects, we find that
the results are quite reliable. The major difference between the two cases concerns the mass
ratio and the total mass of the system (given by $K_{\rm Aa} + K_{\rm Ab}$), which is higher
in the case of full modeling.

Because of the enormous amount of CPU time required to model the deformed (and varying)
shapes of the components with the orbital revolution, it is practically impossible to survey
the whole parameter space and any possible correlations between parameters. Some properties
of the solution and of information contained in the data are as follows:

\begin{itemize}
\item Unlike LCs, the BFs alone do not define both the eccentricity and the orientation of the
      orbit very well. Fixing eccentricity at different reasonable values gives acceptable range
      $e$ = [0.24, 0.33] without any obvious correlation to $\omega$, always being $[112, 116]\degr$.

\item The LC alone defines inclination angle very well because of the small fractional radii of the
      components, but there is a correlation between the radii of the components and the inclination
      angle. The solution shows that {\it both} eclipses are partial - the visible surfaces
      of the component eclipsed in the minima are 35.6\% (primary minimum) and
      40.8\% (secondary minimum).

\item Including asynchronous rotation in the computation of the component's shapes does not
      improve the solution significantly. The presence of fast asynchronous rotation cannot be
      inferred from the photometry alone. The information content of one broad-band LC is
      clearly much smaller than that of many BFs obtained at different phases.

\item By assuming no gravity darkening ($\beta_{\rm Aa} = \beta_{\rm Ab} = 0.00$), the fits
      to the LCs and BFs become poorer by only 5\% and 2\% (according to $\chi_r^2$). This means that
      it is impossible to reliably determine the gravity darkening coefficient from the
      present data.

\item The combined solution shows that the secondary component is slightly hotter than
      the primary. This information cannot be inferred from BF modeling alone.

\end{itemize}

\section{The triple system $\delta$ Velorum}
\label{triple}

The total mass of the whole triple system $\delta$ Velorum as determined by Argyle et al.
(\cite{arg2002}) is rather uncertain: 5.71$^{+1.27}_{-1.08}$ M$_\odot$ (assuming the Hipparcos
parallax, $\pi = 40.90\pm0.38$ mas). The authors give visual magnitudes of the components as
$V_A$ = 1.97 and $V_B$ = 5.55, corresponding to absolute magnitudes of
$M_V (A)$ = 0.02 and $M_V(B)$ = 3.60. The visual orbit was determined based on the data from
1895 until 1999. Because of the most recent periastron passage (2000.8 according to their orbit), any
new positional measurements of the visual pair could significantly improve the orbit.
The new observation (NACO/VLT on April 1, 2008) of Kervella et al. (2009) enables us to measure a separation
between the components of about 0.6\arcsec\ (their Fig.~2), confirming the orbit of Argyle et al. (\cite{arg2002}).

\begin{figure}
\centering
\includegraphics[width=8cm,clip]{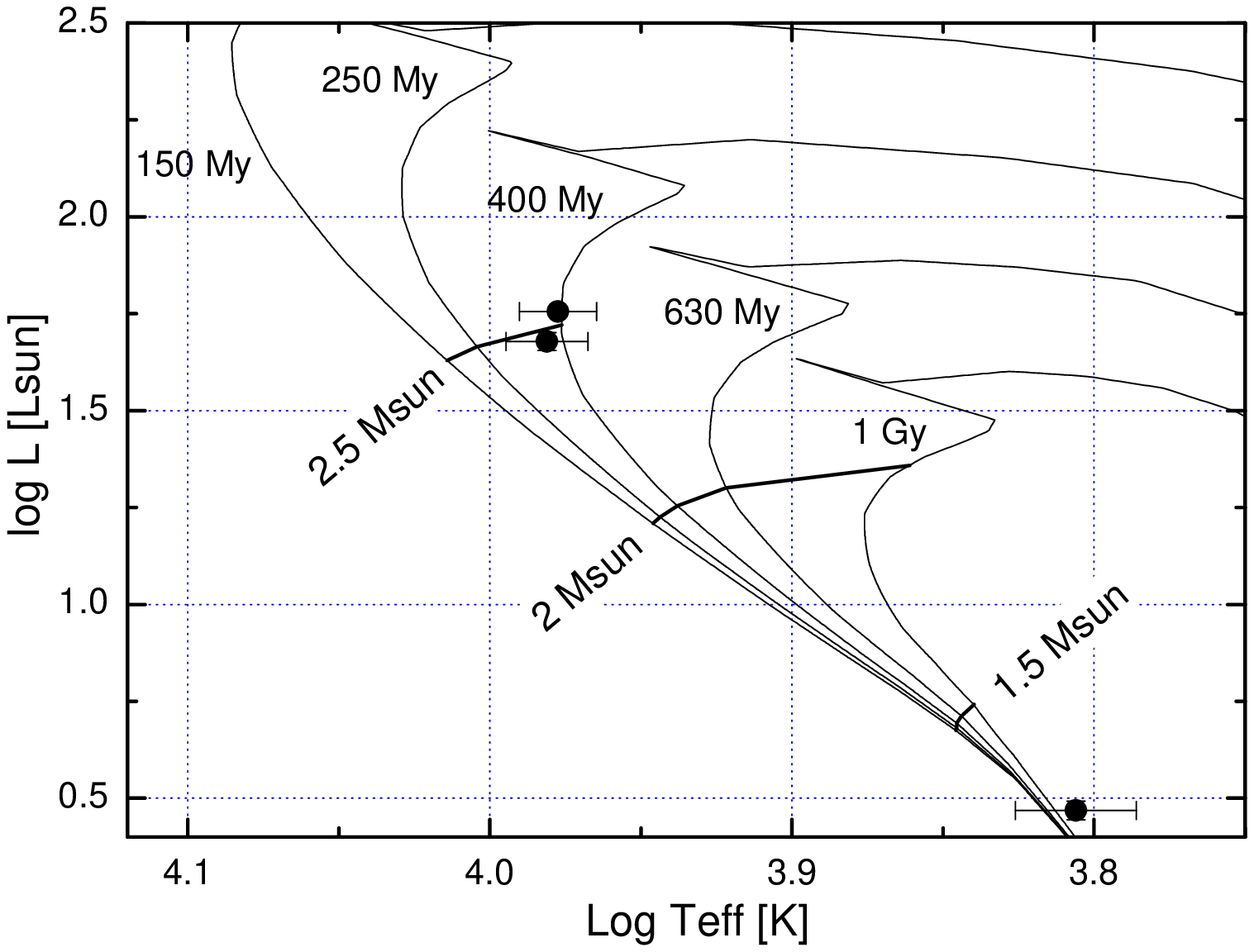}
\caption{Theoretical isochrones (dashed lines) for solar composition (X = 0.70, Y = 0.28, Z = 0.02)
         assuming overshooting adopted from Claret (\cite{clar2004}). Positions of all three components
         of $\delta$ Velorum are shown. Thick solid lines correspond to masses 1.5 M$_\odot$, 2 M$_\odot$,
         and 2.5 M$_\odot$. Luminosities of Aa and Ab correspond to the combined
         solution, while for component B the luminosity was derived from the Hipparcos distance
         and observed visual magnitude. Horizontal error bars for the components of the eclipsing
         pair correspond to the surface temperature ranges due to the gravity brightening.}
\label{isoch}
\end{figure}

Our combined LC and BF solution gives the following luminosities for the components:
$L_{\rm Aa}$ = 56.3 L$_\odot$ and $L_{\rm Ab}$ = 47.1 L$_\odot$. Taking the absolute bolometric magnitude
of Sun as $M^{bol}_\odot$ = 4.75 and bolometric correction for components of $\delta$ Vel (A0V)
as B.C. = $-0.15$ (Popper, \cite{popp1980}), we derive the absolute visual magnitudes of the components
to be $M_V (Aa)$ = +0.519 and $M_V(Ab)$ = +0.719. This gives a combined magnitude of $M_V (A)$ = $-0.138$, which
is substantially brighter than the reliable value determined from the observed visual brightness and Hipparcos
distance by Argyle et al. (\cite{arg2002}). This simple computation, however, does not take into account
that the observer faces the coldest (equatorial) part of the system. The apparent magnitudes of the
eclipsing pair
synthesized by the {\it ROCHE} code (including all proximity effects and gravity darkening),
$U$ = 1.98$\pm$0.02, $B$ = 1.94$\pm$0.02, $V$ = 1.95$\pm$0.02, and $K$ = 1.86$\pm$0.02 are in very good
agreement with the observations: Kervella et al. (\cite{kerv2009}) found $V$ (Aab) = 2.00$\pm$0.02 and
$K$ (Aab) = 1.86$\pm$0.09.

The positions of all three components in the H-R diagram and the theoretical isochrones are plotted
in Fig.~\ref{isoch}. Assuming that the third component is still on the main sequence\footnote{It is the least
massive of the trio, with a significant turn-off from the main sequence at ages older than 1 Gyr, excluded
by the evolutionary state of the eclipsing pair.}, its absolute visual magnitude, $M_V(B)$ = 3.60,
corresponds to F2-F5 spectral type or according to Cox (\cite{cox2000}) to $T^{\rm eff}_{\rm B} = 6100 - 6700$ K
and luminosity $L_{\rm B} = 3.19 - 3.56$ L$_\odot$. In the H-R diagram, component B is slightly above the main
sequence, supporting the earlier limit for the spectral type, namely F2V. The positions of components of the
eclipsing pair Aa, Ab plotted in Fig.~\ref{isoch} were derived from the combined solution to all BFs
and SMEI LC for $T^{\rm eff}_{\rm Aa}$ = 9420~K.

The observed masses and radii of the components correspond most closely to the model predictions (Claret, \cite{clar2004},
$Y$ = 0.70, $Z$ = 0.02, and overshooting) for 400 Myr, when the temperatures of components are assumed to be
$T_{\rm Aa}$ = 9470~K, $T_{\rm Ab}$ = 9370~K, $R_{\rm Aa}$ = 2.643 R$_\odot$, and $R_{\rm Ab}$ = 2.363 R$_\odot$.
The corresponding apsidal motion constants are $\log k_2$ (Aa) = $-2.4972$ and
$\log k_2$ (Ab) = $-2.4626$. The observed radii of stars are still about 5-6\% larger, giving additional
support to measurements of lower temperature and a slightly older age than 400~Myr - soon after 400~Myr,
the temperatures of the components equalize (as observed) because of the slightly faster evolution
of the more massive primary component. The next isochrone available from Claret (\cite{clar2004}) for 630 Myr predicts
that $T_{\rm Aa}$ = 7980~K, $T_{\rm Ab}$ = 8132 K, $R_{\rm Aa}$ = 4.263 R$_\odot$, and
$R_{\rm Ab}$ = 3.393 R$_\odot$. The evolution of the components could, however, be affected by
fast rotation, making the main-sequence phase last longer (see e.g., de Mink, \cite{mink2010}).

\section{Discussion and future work}

Our new data has enabled the first sound analysis of the eclipsing pair Aab in the $\delta$ Velorum system.
The main results are as follows:

\begin{itemize}
\item Both minima are partial. For the given orientation of the orbit and assuming (fairly similar)
      the derived radii of components, the total eclipse and annular transit occur only if $i > 89.8 \degr$.

\item The components of the eclipsing pair are a factor of two smaller than derived interferometrically by
      Kellerer et al. (\cite{kell2007}).

\item The brightness ratio of brightness of components in the SMEI passband (close to the $R$
      passband) is about 0.823. The components are of similar temperature, however, both very
      probably being of spectral type A1V. This is strongly supported by a close agreement with the
      synthetic depth of the primary minimum in the $K$ passband $\Delta$mag I = 0.430, while
      Kervella et al. (\cite{kerv2009}) determined $\Delta$mag I = 0.440$\pm$0.011 from the
      NACO imaging during the primary eclipse.

\item There is no apsidal motion in the system observed during the six-year time-series of the
      SMEI data as indicated by the observed times of the minima.

\item Photocenter motion of the eclipsing pair (corresponding to our solution) is only about
      1 mas: it was not detected by the Hipparcos satellite.

\item Both components rotate very rapidly: the primary at about 56\% of the break-up velocity, the
      secondary at about 62\%. This also means that the components are strongly deformed.
      When the surface can be accurately described by a generalized ROCHE model, the polar flattening
      of the components is 1/10 and 1/12. Because of the fast rotation, the surface temperatures are
      rather non-uniform. The von Zeipel theorem predicts the following ranges:
      9220~K $< T^{\rm eff}_{\rm Aa} <$ 9780~K and 9280~K $< T^{\rm eff}_{\rm Ab} <$
      9880~K\footnote{Surface temperature increases from the equator to the poles.}. This
      effect should be confirmed by long-baseline interferometry.

\item The best fits to the BFs extracted during the primary transit (spectra taken at phases
      0.995, 0.997, 0.000, and 0.002) shows just slight deviations indicating that the
      rotational axis of the primary component is almost perpendicular to the orbital
      plane. The secondary minimum, occurring around phase 0.436, is insufficiently covered
      (a single spectrum at phase 0.441) by the spectroscopic observations to be able to derive
      any information about the rotational axis orientation.

\item The spectral type of $\delta$ Velorum A is most probably A1V, although different approaches
      infer a fairly wide range of results: (i) the BF strength indicates that $T^{\rm eff}_A$ = 9500~K
      (A0V) (see Section~\ref{doppler}); (ii) a combination of the observed visual brightness, Hipparcos
      distance, and radii of the components (simultaneous solution in Section~\ref{simultan})
      supports a similar temperature, 9420~K; (iii) the observed colors $U-V$ = 0.11, $B-V$ = 0.04,
      $V-R$ = 0.05, and $V-I$ = 0.09 imply a later classification, A1-2V (see Stickland \& Hucht,
      \cite{stic1977}); (iv) the shape and intensity of the H$_\beta$ line (in our BESO spectra)
      corresponds most closely to a higher temperature, $T^{\rm eff}_{\rm A}$ = 10000~K and
      $\log g_{\rm A}$= 4.5-4.6 (B9.5V); and (v) Gray (\cite{gray2006}) gives a spectral type of A1Va(n),
      $T^{\rm eff}$ = 9021 K, $\log g$ = 3.79, and $\log$ [M/H] = $-0.33$. Because the system is
      very close, no interstellar reddening is expected. On the other hand, the presence
      of circumstellar material around $\delta$ Velorum cannot be fully excluded.
\end{itemize}

In spite of significant progress, the astrophysical parameters of $\delta$ Vel should still be verified.
The total mass of the system is slightly affected by the unknown nature (multiplicity, rotation rate) of
the third component ($\delta$ Velorum B). To correctly remove its contribution to the observed line
profiles, it will be necessary to acquire its spectrum separately. This task will not be easy: the separation between
the components will become smaller until 2013 when it reaches 0.38\arcsec. The separation then
should slowly increase until 2067, reaching about 2.9\arcsec.

Our phase-resolved spectra may still need to be precessed by the Fourier disentangling (see Hadrava,
\cite{hadr1995}) to obtain the individual spectra of components and their astrophysical parameters
($\log g$, $T^{\rm eff}$, metallicity etc.). Because of the strong rotational mixing, any chemical
peculiarity is, however, excluded.
Most importantly, the combination of the present data with long-baseline interferometry will be
required, which may possibly detect the flattening of the components and "polar caps" brightening caused
by the rapid rotation. This additional dataset would break the parameter correlations complicating the
present modeling.

Additional high-resolution spectroscopy should be secured during the eclipses to study the
Rossiter-McLaughlin effect. Photometric eclipse modeling can still be improved by dedicated
multi-color photometry of the transits, which is important for a clearer definition of the
surface brightness ratio.

Another possible way to improve the outer visual orbit is to use timing of the minima of
the eclipsing pair, which should show a light-time effect. The expected semi-amplitude, for the visual
orbit of Argyle et al. (\cite{arg2002}), is about 1.5 hour. In view of the large $v \sin i$ of the
components and the blending of their profiles, systemic-velocity changes in the eclipsing pair
would be hardly detectable nor useful.

\begin{acknowledgements}
M.V. and T.P. acknowledge support from the EU in the FP6 MC ToK project MTKD-CT-2006-042514.
This work has partially been supported by VEGA project 2/0038/10.

This publication is supported as a project of the  Nordrhein-Westf\"alische Akademie
der Wissenschaften und der K\"unste in the framework of the academy program by the Federal
Republic of Germany and the state Nordrhein-Westfalen.

M.A. acknowledges research funding granted by the {\it Deutsche Forschungsgemeinschaft} (DFG)
under the project RE 1664/4-1. A.B. acknowledges support from DFG in program NE 515/32-1.

The research made use of the SIMBAD database, operated at the CDS, Strasbourg, France.

\end{acknowledgements}

\newpage

\begin{figure*}
\centering
\includegraphics[width=16cm,clip]{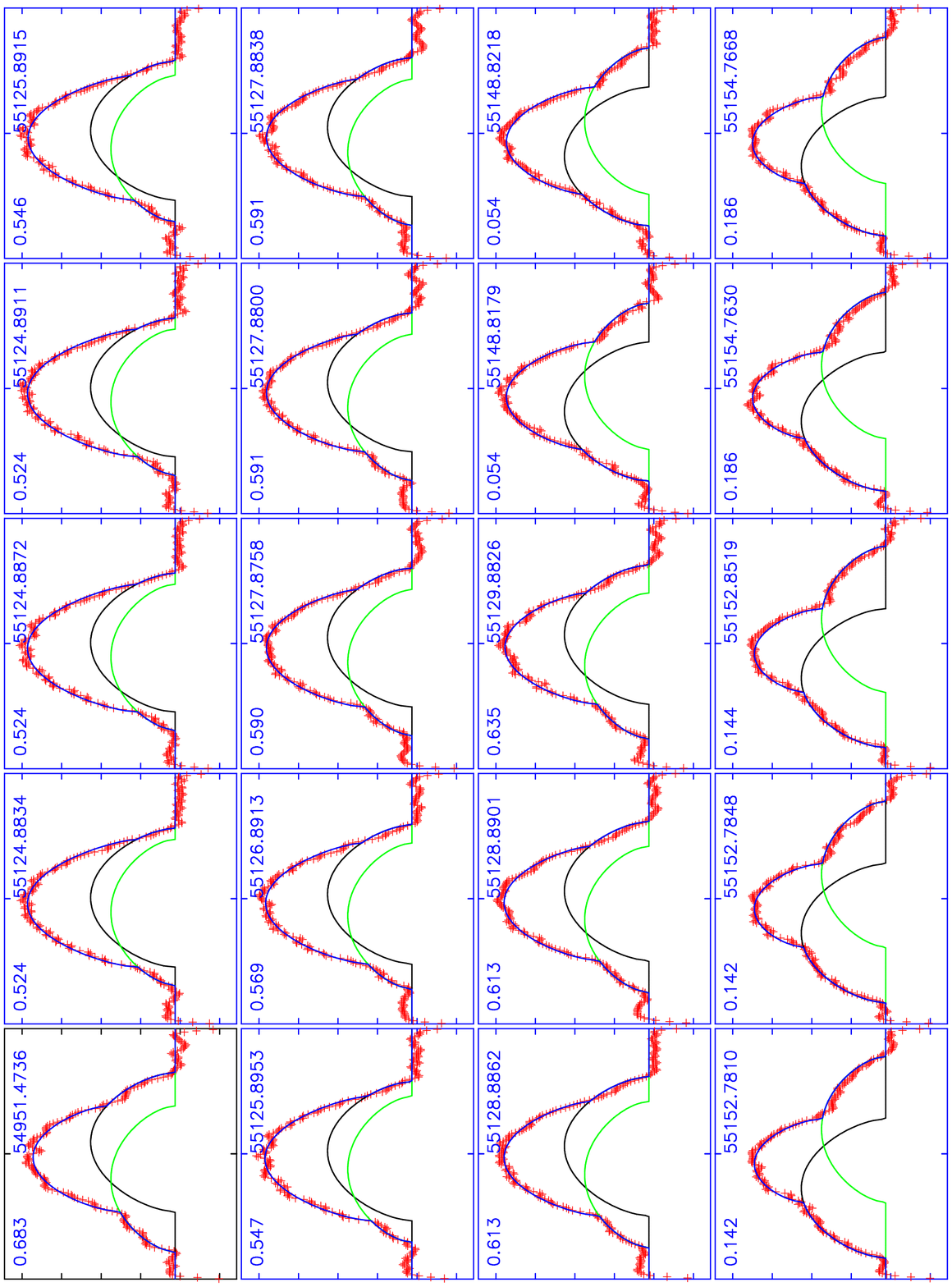}
\caption{Rotational profile fit to all broadening functions outside eclipses. The panels give
the orbital phase and the heliocentric Julian date (with 2\,400\,000 subtracted). The profile
of the primary component (Aa) is plotted in black, the profile of the secondary component
(Ab) in green, and the combined profile in blue.}
\end{figure*}

\begin{figure*}
\addtocounter{figure}{-1}
\centering
\includegraphics[width=16cm,clip]{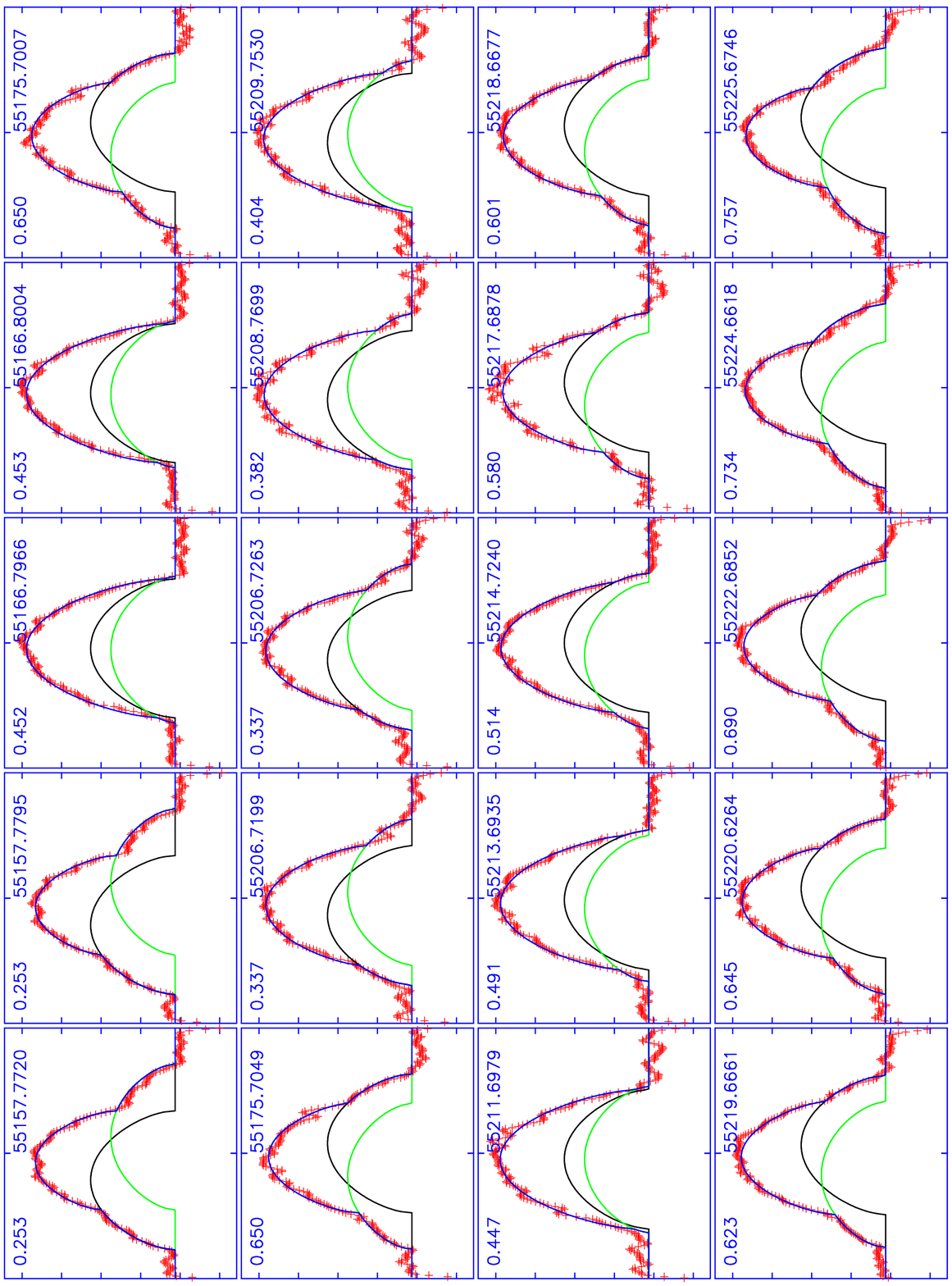}
\caption{(continued)}
\end{figure*}

\begin{figure*}
\addtocounter{figure}{-1}
\centering
\includegraphics[width=16cm,clip]{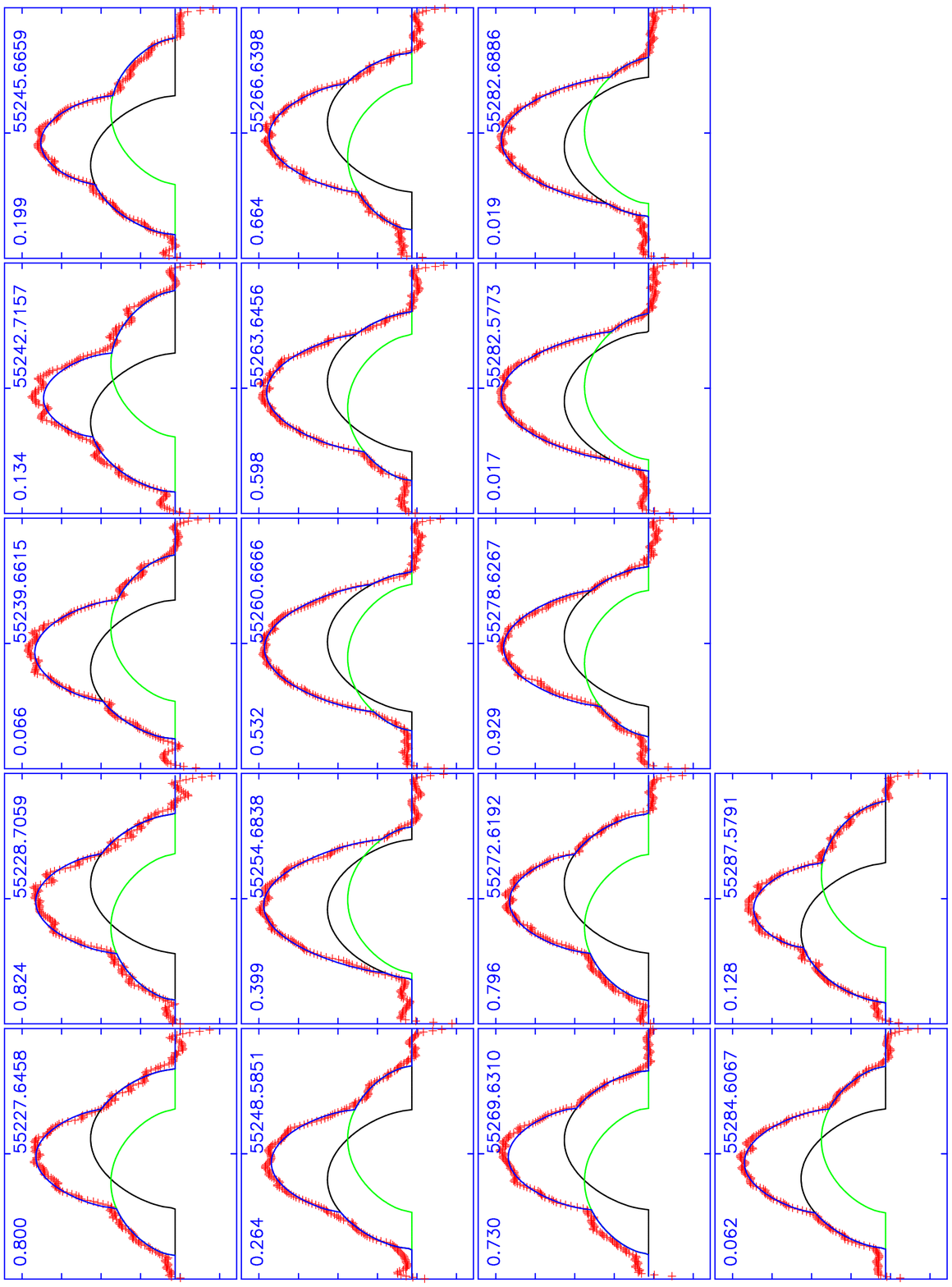}
\caption{(continued)}
\end{figure*}

\begin{figure*}
\centering
\includegraphics[width=16cm,clip]{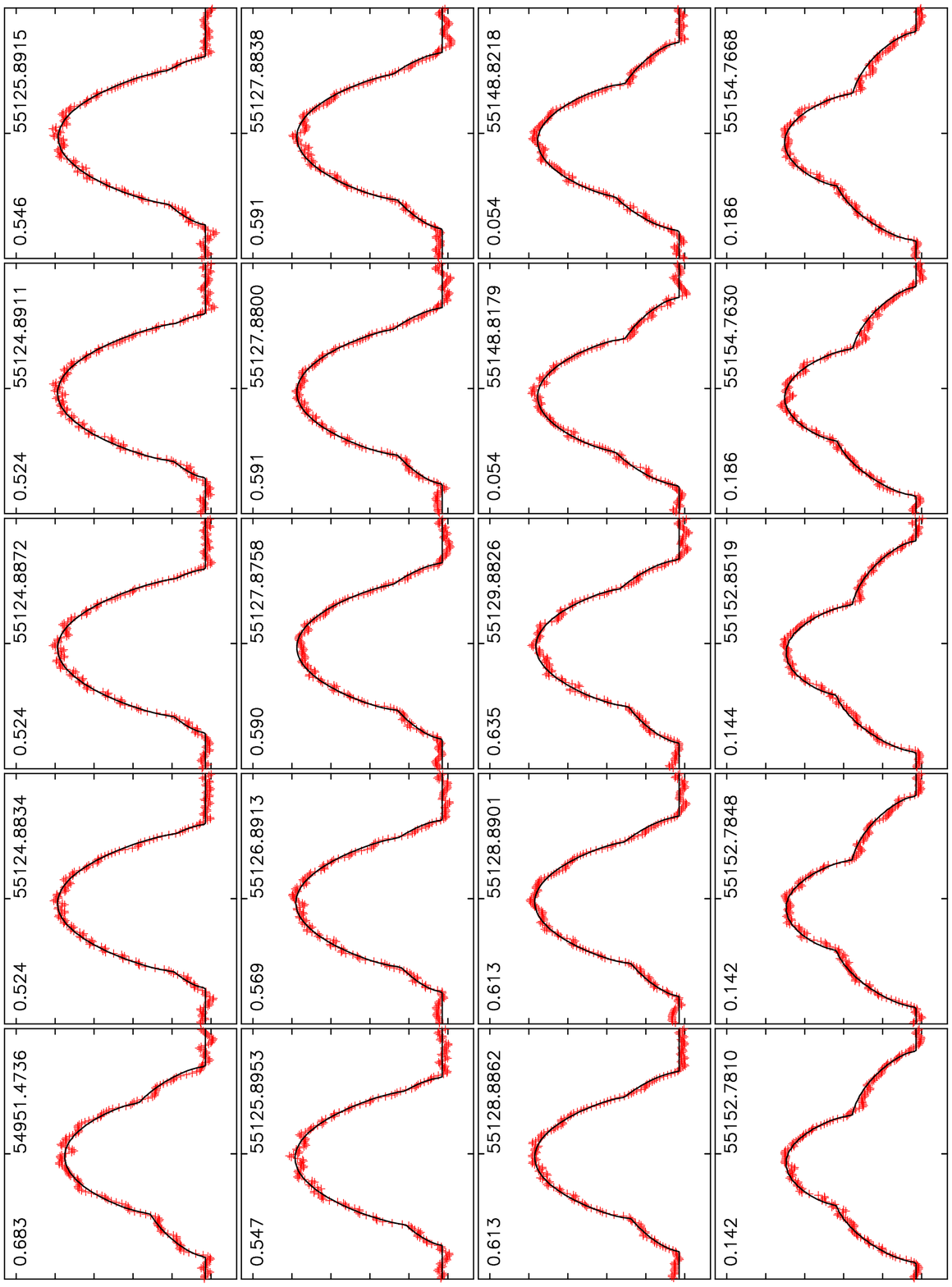}
\caption{The ROCHE-model fit to all broadening functions. The panels give
the orbital phase and the heliocentric Julian date (with 2\,400\,000 subtracted)}
\end{figure*}

\begin{figure*}
\addtocounter{figure}{-1}
\centering
\includegraphics[width=16cm,clip]{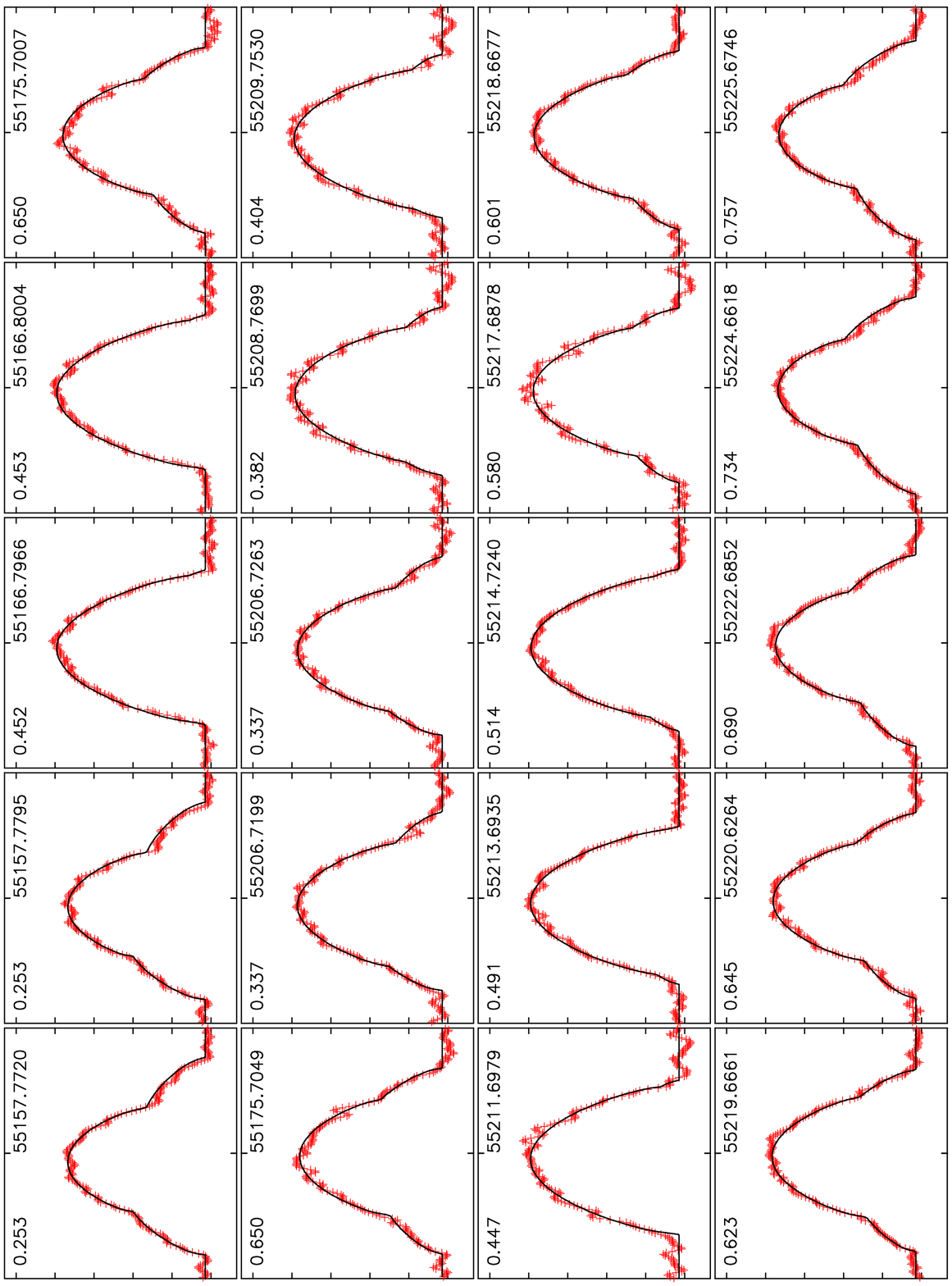}
\caption{(continued)}
\end{figure*}

\begin{figure*}
\addtocounter{figure}{-1}
\centering
\includegraphics[width=16cm,clip]{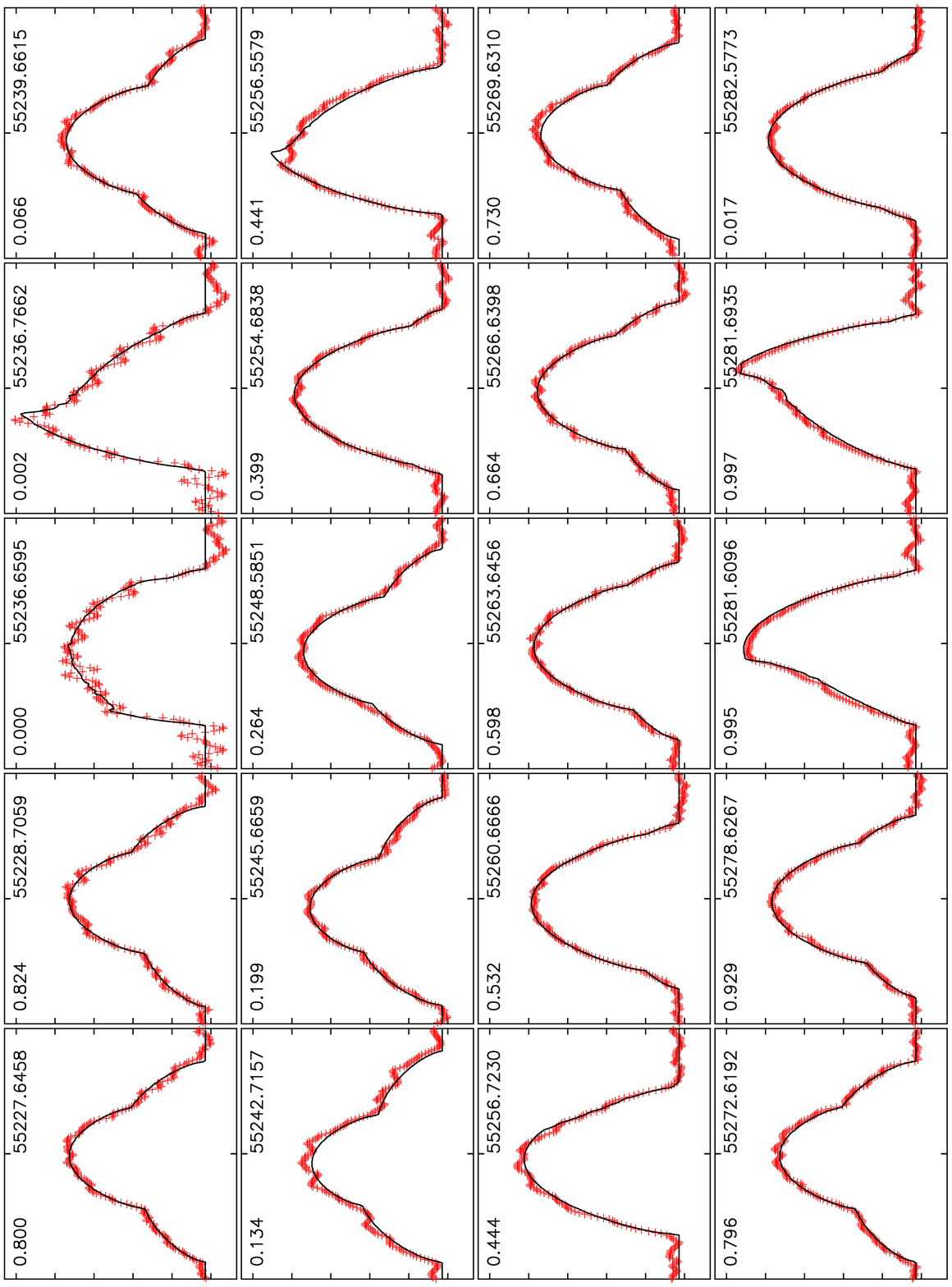}
\caption{(continued)}
\end{figure*}

\begin{figure*}
\addtocounter{figure}{-1}
\centering
\includegraphics[width=4cm,clip]{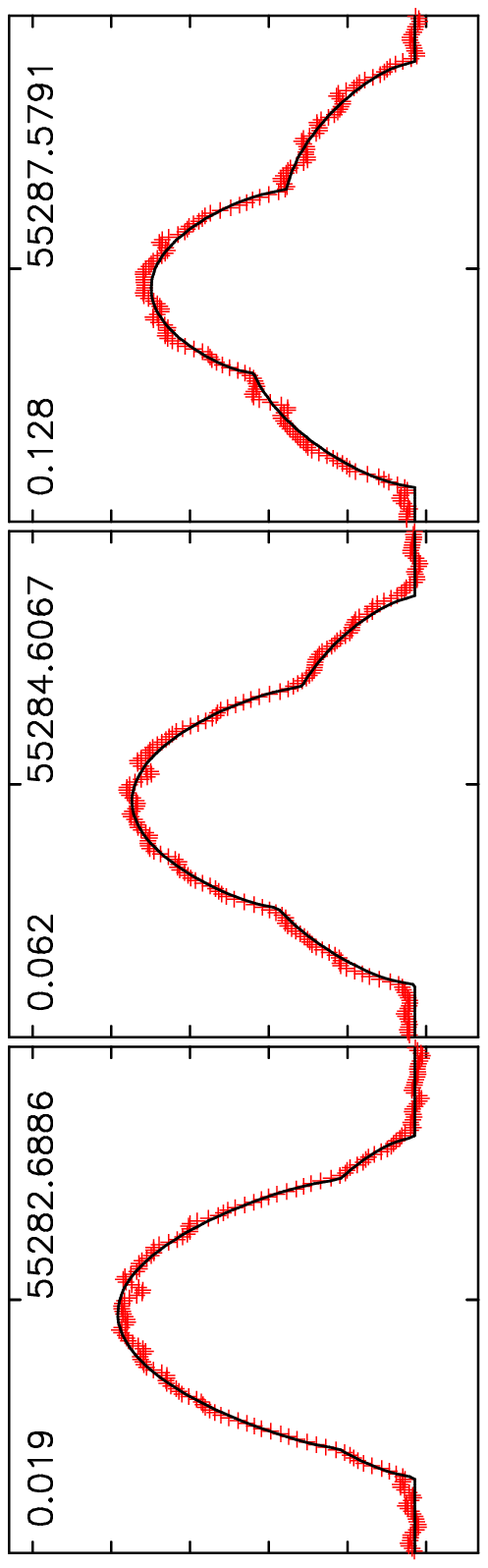}
\caption{(continued)}
\end{figure*}

\end{document}